\documentstyle[epsf, twocolumn,seceq]{jpsj}
\title
{Non-Fermi Liquid Behavior in Dilute Quadrupolar System Pr$_{x}$La$_{1-x}$Pb$_3$ with $x$$\le$0.05}
\author
{Tatsuya {\sc KAWAE} \footnote{E-mail : kawaetap@mbox.nc.kyushu-u.ac.jp},
 Tatsuharu {\sc YAMAMOTO},
 Kenji {\sc YURUE},
Naoyuki {\sc TATEIWA},
 Kazuyoshi {\sc TAKEDA}
and
Tetsuo {\sc KITAI}$^1$ 
 }
\inst{
Department of Applied Quantum Physics, Faculty of Engineering, Kyushu
University, Fukuoka 812-8581\\
$^1$Faculty of Engineering, Kyushu Institute of Technology, Kitakyushu 804-8550\\
}
\recdate
{
\today
}
\abst
{ We have studied the low-temperature properties of Pr$_{x}$La$_{1-x}$Pb$_{3}$ with non-Kramers $\Gamma _{3}$ quadrupolar moments of the crystal-electric-field ground state, for a wide concentration range of Pr ions. For $x$$\le$0.05, the specific heat $C/T$ increases   monotonically below $T$=1.5 K, which can be scaled with a characteristic temperature $T^{*}$ defined at each concentration $x$. The electrical resistivity $\rho$$(T)$ in the corresponding temperature region shows a marked decrease deviating from a Fermi-liquid behavior $\rho$$(T)$$\propto$$T^{2}$. The Kondo effect arising from the correlation between the dilute $\Gamma _{3}$ moments and the conduction electrons may give rise to such anomalous behavior.
}
\kword
{non-Fermi liquid, Kondo effect, $\Gamma_{3}$ quadrupolar moments, PrPb$_{3}$, specific heat
}
\begin{document}
\sloppy
\maketitle
%
%
%
During the past decade, several U and Ce-based compounds have been reported to show a non-Fermi-liquid (NFL) behavior with the unusual temperature dependence of the electronic specific heat and the magnetic susceptibility, which are clearly inconsistent with the Fermi liquid theory. Two quite different scenarios have been proposed.
(i) The competition between the Kondo effect and the RKKY interaction for the $f$-electrons is thought to play an important role in the NFL behavior. Hence it has been observed near the quantum critical point where the ground state changes from an antiferromagnetic ordering or a spin glass state to a nonmagnetic one~\cite{rf:1,rf:2}. (ii) Contrary to the case mentioned above, a single-ion multichannel Kondo effect arising from the overscreening of the spin $S$ of the 4$f$- or 5$f$-electrons due to the $N$ scattering channels of the conduction electrons, with $N>2S$, is proposed to lead to the NFL behavior.  Cox showed that the two-channel Kondo effect can be realized in U$^{4+}$($5f^{2}$) ions at a site of cubic symmetry with the crystal-electric-field  (CEF) ground state of  a non-Kramers $\Gamma _{3}$ doublet, that is referred to as the quadrupolar Kondo effect~\cite{rf:3}.  The $\Gamma _{3}$ doublet  has no magnetic moments, but it has electric quadrupolar moments which can be described by pseudospin moments with $S=1/2$. Thus the correlation between  the $\Gamma _{3}$ doublet  and the charge of the conduction electrons  involves two electron channels, giving an overcompensation of the $\Gamma _{3}$ quadrupolar moment by conduction electrons below a characteristic temperature $T_{K}$ in the quadrupolar Kondo effect. This leads to the NFL behavior such as  ln$T$-dependence of the specific heat and  the quadrupolar susceptibility, in addition to  $\sqrt{T}$-dependence of the electrical resistivity~\cite{rf:3,rf:4}. Moreover,  the residual entropy in the ground state is 1/2$R$ln2.

 Seaman {\it et al.} reported that Y$_{0.8}$U$_{0.2}$Pd$_{3}$ shows an NFL behavior,  and ascribed the results to the quadrupolar Kondo effect~\cite{rf:5}.  After their study, a number of U-based compounds have been reported to show an NFL behavior~\cite{rf:6,rf:7}.  However, there remain several questions to understand the NFL behavior in these systems within the framework of  the quadrupolar or two-channel Kondo effect~\cite{rf:8,rf:9}.  

  Pr-based compounds are  good candidates to study the quadrupolar Kondo effect, since the CEF level scheme of Pr$^{3+}$ is the same as that of U$^{4+}$ as long as it is in the same crystal field symmetry. The study focusing on the fluctuation of the $\Gamma _{3}$ quadrupolar moments  in Pr-based compounds  was carried out in PrInAg$_{2}$, for the first time~\cite{rf:10}.  It showed no quadrupolar ordering (QPO) down to 50 mK, but a $T$-linear variation of the specific heat with a very large value of $C/T$ ($\sim $ 6.5 J/mol K$^{2}$), which is different from the ln$T$-dependence. Recently, superconductivity was discovered in PrOs$_{4}$Sb$_{12}$, in which the pairing interaction was discussed in terms of the fluctuation of the $\Gamma _{3}$ moments~\cite{rf:11}. However, clear experimental evidence on the correlation between the $\Gamma _{3}$ moments and the conduction electrons has not been obtained so far. 

We have studied the low-temperature properties of Pr$_{x}$La$_{1-x}$Pb$_{3}$ with a non-Kramers $\Gamma _{3}$ doublet in the CEF ground state by changing $x$ to clarify how the fluctuation of the $\Gamma _{3}$ moments is suppressed at low temperatures~\cite{rf:12}.  In this paper, we report the observation of NFL behavior in $C/T$ and the electrical resistivity in a very dilute region of the $\Gamma _{3}$ moments for $x$$\le$0.05, which is likely ascribed to the correlation between the $\Gamma _{3}$ moments and the conduction electrons. 

The magnetic properties of the host system PrPb$_{3}$ have been studied extensively. The ground state is a $\Gamma _{3}$ doublet  and the first excited state is a $\Gamma _{4}$ triplet  with an energy difference of 19 K from the ground state~\cite{rf:13}. Antiferro-QPO appears  at $T=$0.4 K. Both PrPb$_{3}$ and LaPb$_{3}$ have the AuCu$_{3}$-type cubic structure with the lattice parameter $a=$ 4.867 {\AA} and 4.903 {\AA}, respectively. Thus La ions can be substituted for Pr ions without a change in crystal symmetry for any concentration. 

The samples of Pr$_{x}$La$_{1-x}$Pb$_{3}$ are prepared by the Bridgeman method. From the powder X-ray diffraction pattern, the crystal structure is confirmed to be cubic. The magnetization and the magnetic susceptibility are measured by a Quantum Design SQUID magnetometer. The specific heat is measured by a heat pulse method, and the electrical resistivity by a Linear Research LR700 four-wire ac resistance bridge, using polycrystalline samples.

The main results for the Pr concentration of 0.2$\le$$x$$\le$1 were reported in a previous paper~\cite{rf:12}. For a wide range of Pr concentration, the QPO is absent, and instead the low-temperature specific heat shows a $T$-linear variation with a large coefficient, similarly to PrInAg$_{2}$, which can be reproduced by the model for amorphous materials with a random configuration of two-level system (RTLS) with the maximum energy splitting $E_0$/$k_B$$\sim$3 K.  This implies that the fluctuation of the  $\Gamma _{3}$ moments is suppressed by the formation of RTLS. The ground state with the $\Gamma _{3}$ doublet in the CEF level scheme does not change for the Pr concentration down to $x$=0.2. This is consistent with the fact that $E_0$/$k_B$, which reflects the maximum distortion of the CEF due to La ions, is much smaller than the energy splitting between the CEF ground state and the first excited state in PrPb$_{3}$. 

The temperature dependence of the magnetic susceptibility for the Pr dilute region ($x$$\le$0.05) is qualitatively the same as that for the Pr concentrated region ($x$$\ge$0.1).  Figure 1 shows the temperature dependence of the magnetic susceptibility for $x$=0.05 measured at $H$=0.3 T. The susceptibility is well reproduced by the CEF level scheme with the $\Gamma _{3}$ doublet in the ground state, as described by a solid line in Fig. 1, such as the case for $x$$\ge$0.2.  The Van Vleck susceptibility is in good agreement with the results obtained by Bucher {\it et al.}~\cite{rf:14}.  The increase in the susceptibility below 5 K is considered to come from a small modification of the CEF levels due to the La ions, discussed in a previous paper~\cite{rf:12}. In addition, there may be a contribution from the correlation between the $\Gamma _{3}$ moments and the conduction electrons~\cite{rf:15}, which will be discussed later. The magnetization curve at 2 K is shown in the inset. It indicates  that  the ground state of the CEF level does not change up to 5 T.  
\begin{figure}
\begin{center}
\epsfxsize=8cm
\epsfbox{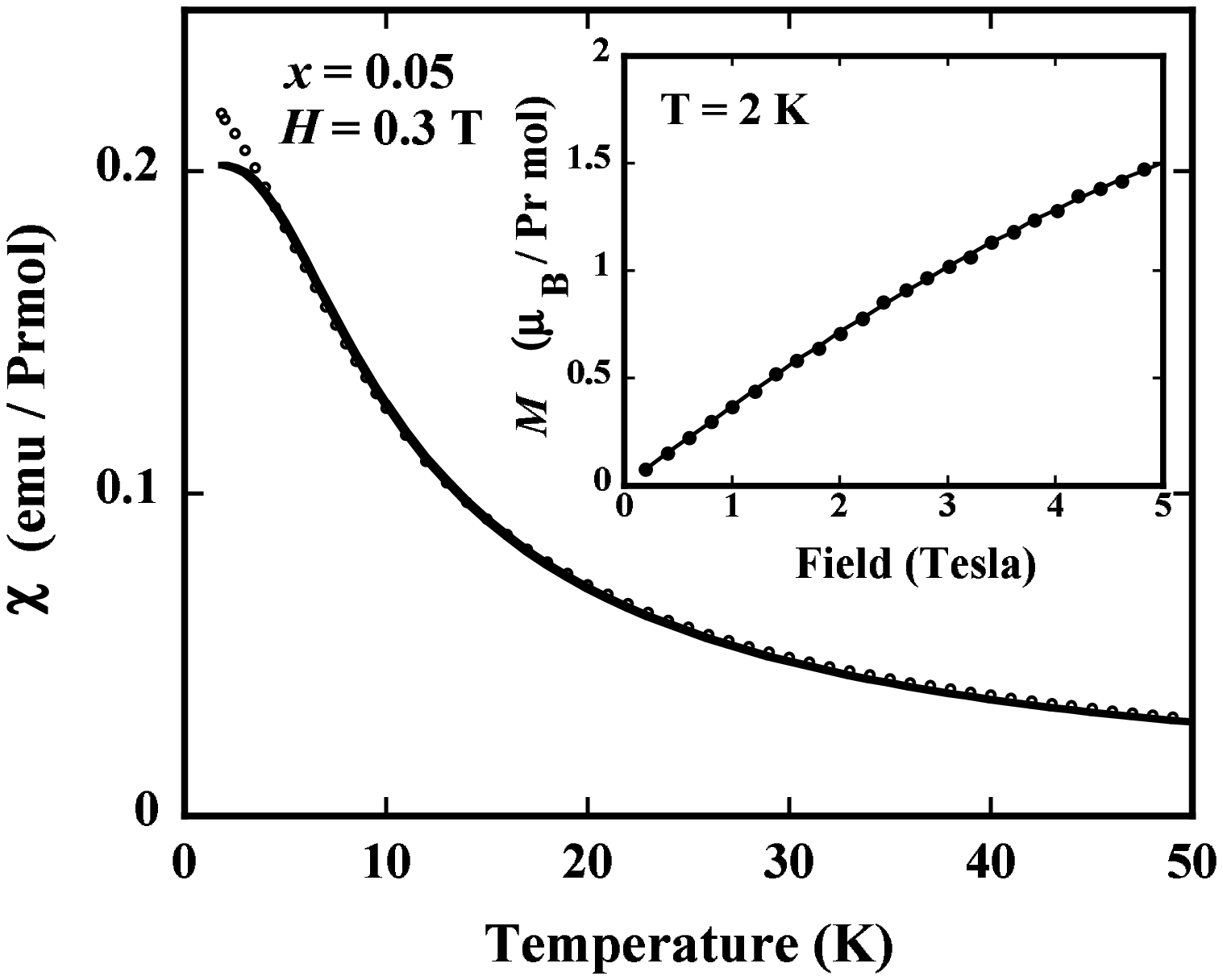}
\caption{Temperature dependence of the susceptibility for $x=0.05$. The solid line represents the calculation based on the CEF level scheme. The ground state is $\Gamma_{3}$ and the first excited state is $\Gamma_{4}$ with the energy splitting of 8.7 K. The next state is $\Gamma_{5}$ with the energy splitting of 14.4 K from the ground state. Inset: The magnetization curve for $x=0.05$ at 2  K. The solid line is a guide to the eye.
}
\label{fig:1}
\end{center}
\end{figure}

The temperature dependence of the specific heat in Pr$_{x}$La$_{1-x}$Pb$_{3}$ for $x$$\le$0.1 is shown in Fig. 2(a), together with the data for the reference compound LaPb$_{3}$.  A clear anomaly  due to a superconducting transition of LaPb$_{3}$ is seen in each curve. The superconductivity is suppressed with increasing Pr concentration as shown in the inset of Fig. 2(b). The concentration dependence of $T_{C}$ in the present experiment is in good agreement with previous results,~\cite{rf:14} and implies that Pr$^{3+}$  ions are in the  nonmagnetic $\Gamma _{3}$ state. 
\begin{figure}
\begin{center}
\epsfxsize=8cm
\epsfbox{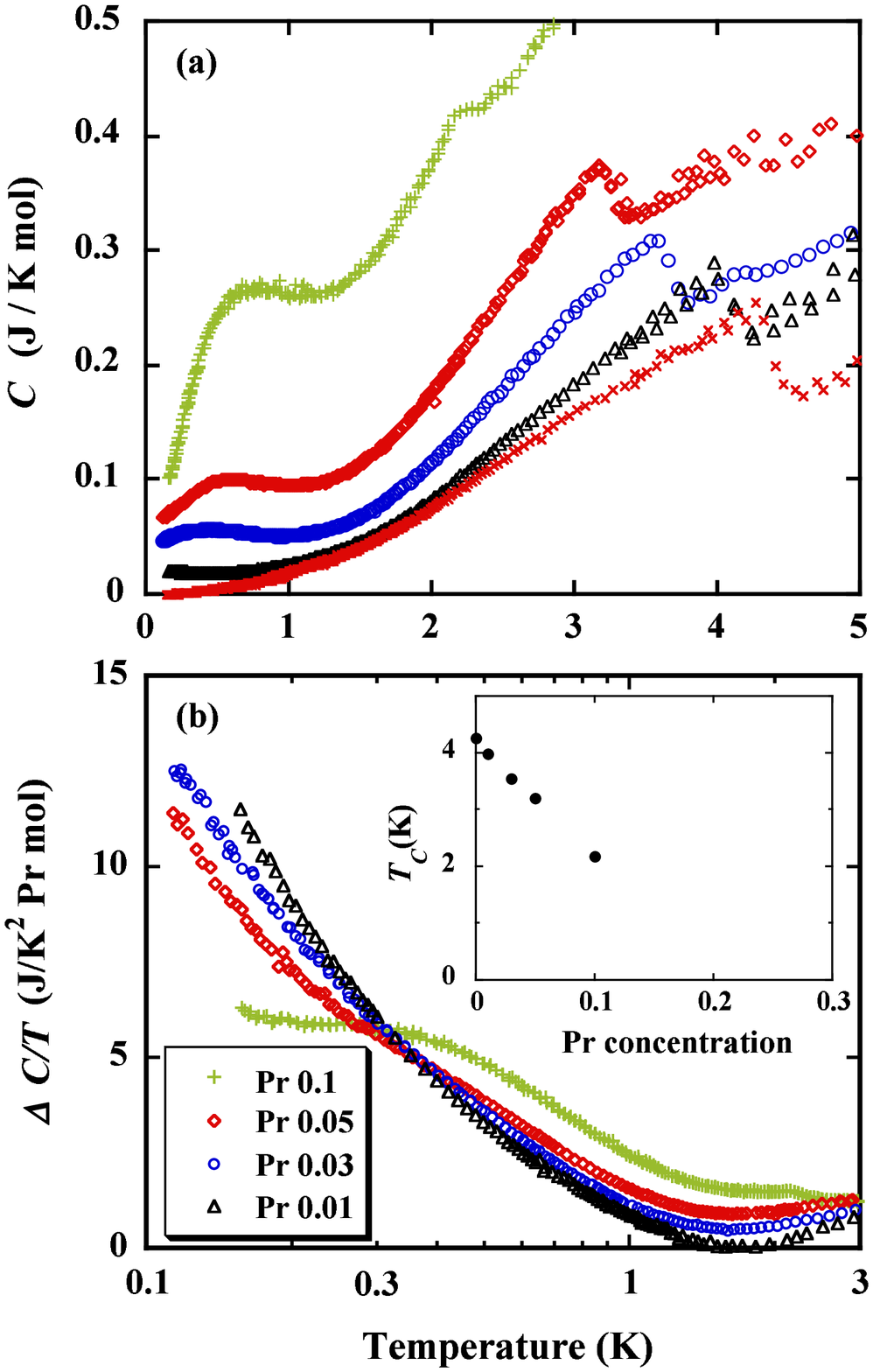}
\caption{(a) Temperature dependence of the specific heat. From top to bottom: $x=$ 0.1, 0.05, 0.03, 0.01, and a reference compound LaPb$_{3}$.   (b) $\Delta$$C/T$ plotted on a logarithmic temperature scale for $x=$ 0.1, 0.05, 0.03, and 0.01, where the specific heat  is normalized by the Pr concentrations and the background contribution estimated from the specific heat of LaPb$_3$ below 4 K is subtracted. Inset: 
 Pr concentration dependence of the superconducting transition temperature of LaPb$_{3}$. 
}
\label{fig:2}
\end{center}
\end{figure}

Below 1.5 K, one can see a distinct hump of the specific heat in Pr$_{x}$La$_{1-x}$Pb$_{3}$, which grows with the Pr concentration.
In Fig. 2(b), we show $\Delta$$C/T$ plots on the logarithmic scale of temperature, where the background contribution is estimated from the specific heat of LaPb$_3$, and is subtracted from that of  Pr$_{x}$La$_{1-x}$Pb$_{3}$. This estimation does not take into account the correction from the superconducting electrons.  Thus  a small anomaly is observed, around 2.2 K for  $x=$0.1, for example.  In the following discussion below 1.5 K, however, the specific heat due to the superconducting electrons is considered to be negligible, because it is much smaller than the additional specific heat due to Pr ions. 

$\Delta$$C/T$ for $x=$0.1 is in agreement with the curve estimated by the RTLS model with the maximum energy splitting $E_0$/$k_B$$=$2.3 K, as in the case for the Pr concentration of 0.2$\le$$x$$\le$0.9. On the other hand, $\Delta$$C/T$ for $x$$\le$0.05 increases monotonically below 1.5 K and cannot be reproduced by the RTLS, which is entirely different from that for $x$$\ge$0.1. 
 For $x$=0.05, a slight hump is seen around 0.4 K, which can be ascribed to traces of the RTLS. In contrast,  a small upturn below 0.2 K for $x$=0.1 can be caused by the same origin as that for $x$$\le$0.05. 
 It is apparent that anomalous $\Delta$$C/T$ does not come from impurity ions or magnetic ordering of Pr ions, because  the temperature dependence of the anomalies is not so sensitive to  the concentration. A Schottky specific heat from the splitting of the Pr nuclear spin state due to the hyperfine magnetic field is also ruled out, because the temperature dependence of the anomaly is clearly different from the Schottky one.   
\begin{figure}
\begin{center}
\epsfxsize=8cm
\epsfbox{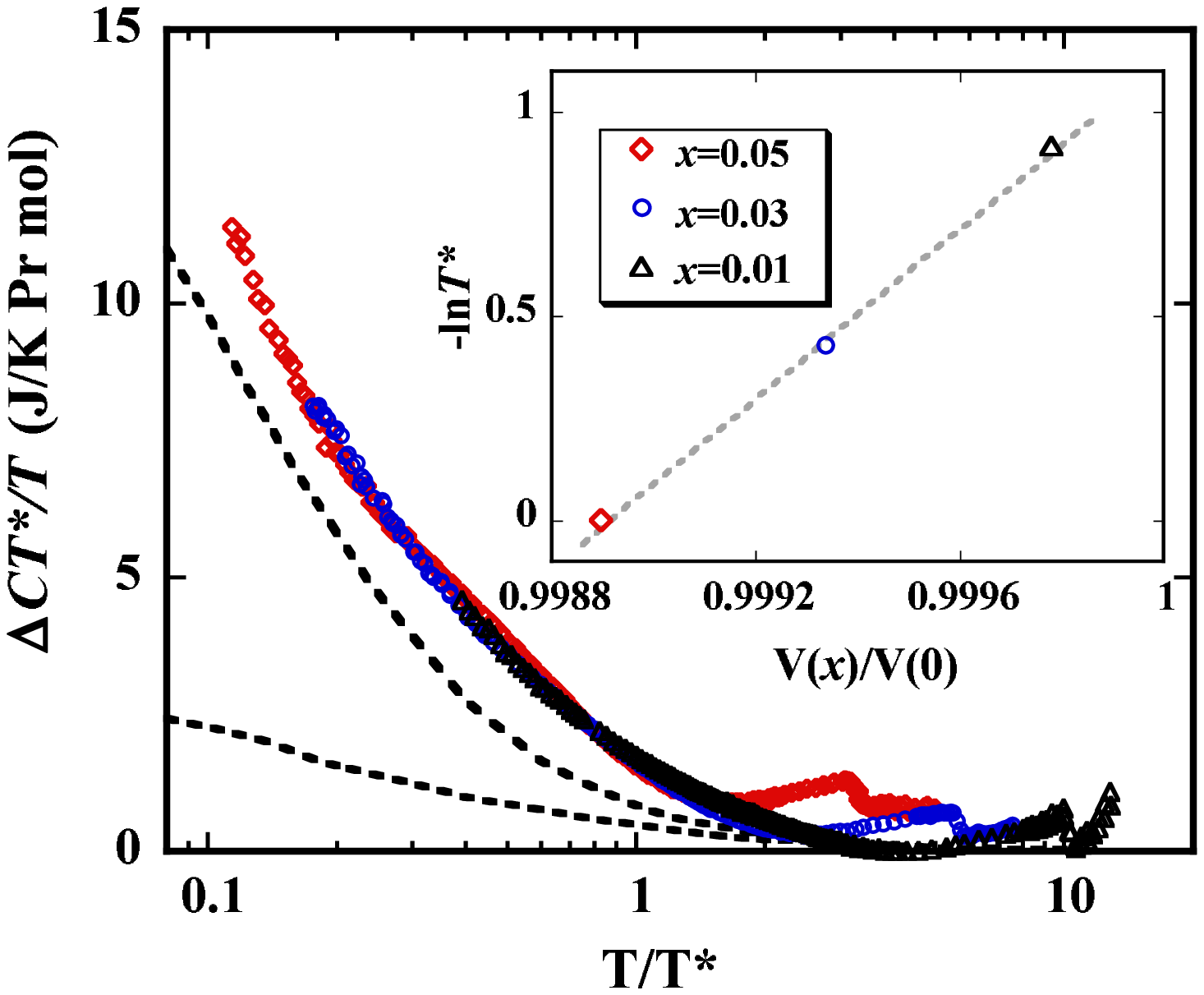}
\caption{ $\Delta$$CT^{*}/T$ versus log$(T/T^{*})$. $T^{*}$ is taken to be 1 K for $x$=0.05 (red), 0.65 K for $x$=0.03 (blue) and 0.40 K for $x$=0.01 (black). The lower dashed line is the theoretical calculation for $T_{K}/T^{*}$=2 in the two-channel Kondo theory and the upper dashed line that for $\mu_{0}H/k_{B}T_{K}$=1 with $T_{K}/T^{*}$=2  obtained from ref. 16. Inset: $V(x)/V(0)$ versus ln$T^{*}$. The volume $V(x)$ is estimated from the lattice parameter. The dashed line is a linear fit to the data.
}
\label{fig:3}
\end{center}
\end{figure}
  
It was found that the data is scaled with a characteristic temperature $T^{*}$ as shown in  Fig. 3, where $T^{*}$ is taken to be 1 K for $x$=0.05, 0.65 K for $x$=0.03 and 0.40 K for $x$=0.01. This indicates that the NFL behavior with the enhancement in $\Delta$$C/T$ for $x$$\le$0.05 is understood as a single-ion effect of the $\Gamma _{3}$ moment and does not come from the superpose of the Schottky specific heat due to the splitting of the $\Gamma _{3}$ moments. $T^{*}$ changes exponentially as shown in the inset of Fig. 3, which suggests that the increase in $T^{*}$ with the concentration $x$ is explained by the increase in hybridization width due to the compression of the lattice by increasing the concentration $x$. 

Let us compare the present experimental results with the theoretical calculation from the $S=$1/2 two-channel Kondo system~\cite{rf:16}, because the scaling behavior of $\Delta$$C/T$ may be related to the quadrupolar Kondo effect. The data, however, cannot be reproduced by the two-channel Kondo theory as plotted by the lower dashed line in Fig. 3, where the Kondo temperature $T_{K}$ is assumed to be $T_{K}/T^{*}$=2.  The entropy change for $x$$\le$0.05 from 0.1 K to 2 K is about 0.8$R$ln2.  Already at 2 K, it exceeds 1/2$R$ln2 which is the theoretical prediction. 
 
 The CEF level of the Pr ions is distorted by the La ions as shown for $x$$\ge$0.1. This means that the La ions induce the strain fields at the Pr sites. The strain field in the quadrupolar system corresponds to the magnetic field in the spin system~\cite{rf:17}. Thus the present results should be compared with the two-channel Kondo theory under strong magnetic fields.  The maximum splitting $E_{0}/k_{B}$=2$\sim$3 K suggests that the strain field, i.e., the magnetic field, is of the same order as the Kondo temperature.
 The upper dashed line in Fig. 3 is obtained by assuming $T_{K}/T^{*}$=2 for $\mu_{0}H/k_{B}T_{K}$=1  in the theoretical calculation~\cite{rf:16}, which indicates that the data may be reproduced by choosing optimum fields and $T_{K}$. 

We measured the electrical resistivity of Pr$_{x}$La$_{1-x}$Pb$_3$ to investigate the scattering of the conduction electrons by the $\Gamma _{3}$ moments.  The temperature dependence of the resistivity $\rho$$(T)$ higher than 10 K can be explained by the scattering between the conduction electrons and phonons and/or CEF excitations, and gives a residual resistivity ratio $\rho$(300 K)/$\rho$(10 K)$\sim$40. Figure 4(a) shows the magnetic field dependence of $\rho$$(T)$ for $x$=0.05. At $H$=0 T, the sudden drop at $\sim$7 K is due to a superconducting transition of the Pb phase, and the drop at $\sim$4 K is due to a superconducting transition of LaPb$_{3}$.  At $H$$\ge$0.2 T, the two-step transition disappears. At higher temperatures, $\rho$$(T)$ follows the Fermi-liquid relation $\rho$$(T)$=$\rho$$_{0}$+$AT^{2}$ with $A$=0.0018$\pm$0.0002 ($\mu$$\Omega$cm/K$^{2}$), whereas it deviates from a $T^2$ dependence  at a certain temperature $T_{D}$ and decreases markedly with lowering temperature.  $T_{D}$, e.g., 4.5 K at $H$=1 T and 2 K at $H$=4 T in Fig. 4(a), is apparently  higher than the superconducting transition temperature of Pr$_{0.05}$La$_{0.95}$Pb$_{3}$ in the respective magnetic fields expected from the susceptibility and specific heat measurements. This demonstrates that the decrease does not come from the superconductivity of LaPb$_{3}$. Similar features of the marked decrease in $\rho$$(T)$ are observed for $x$=0.03~\cite{rf:18}.  $\rho$$(T)$ in the lower fields becomes constant or $\sim$0 ($\mu$$\Omega$cm) at the lower temperatures, which is probably ascribed to the superconductivity of LaPb$_{3}$. 
\begin{figure}
\begin{center}
\epsfxsize=8cm
\epsfbox{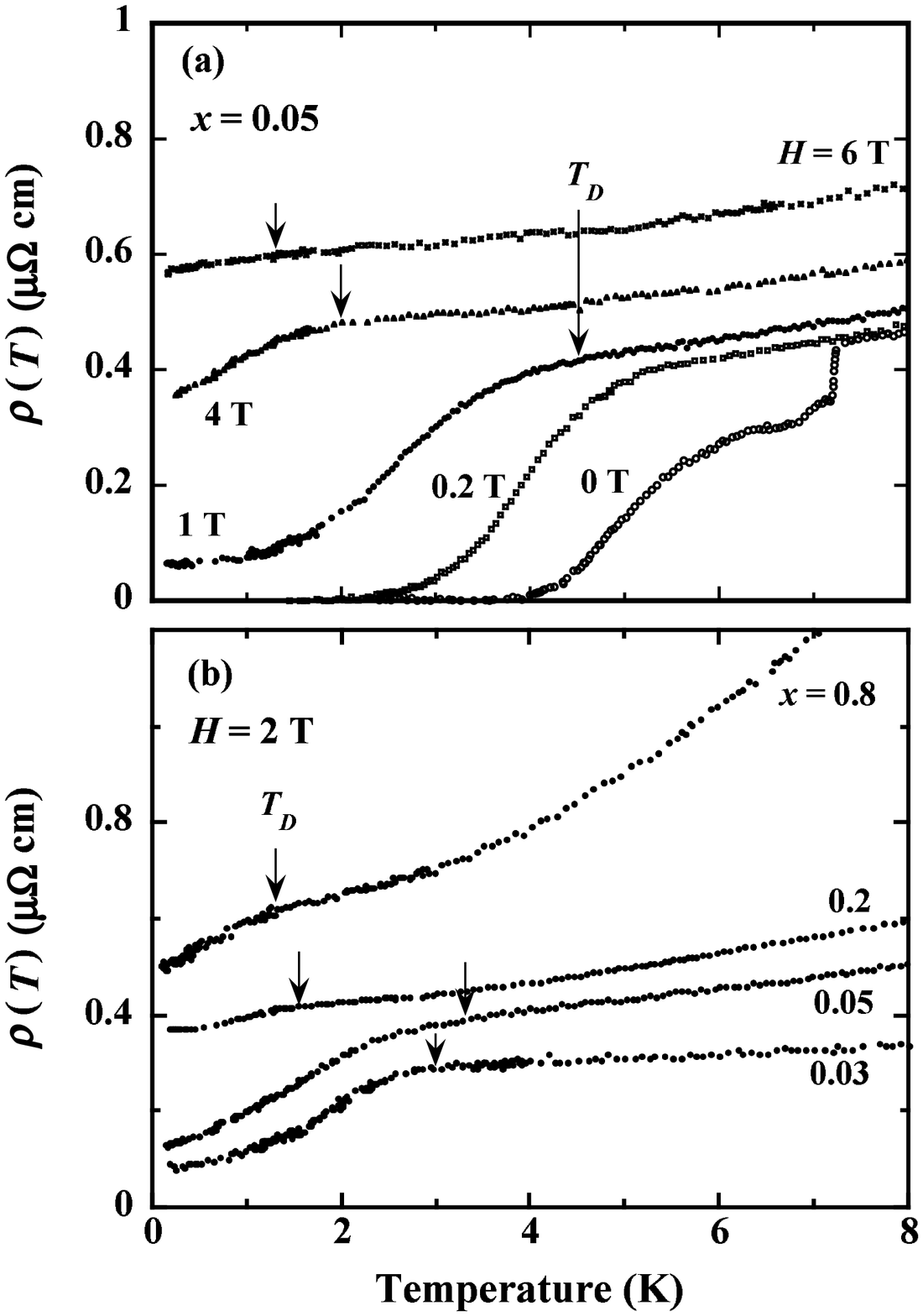}
\caption{(a) $\rho$$(T)$ at $H$=0 T, 0.2 T, 1 T, 4 T,  and 6 T. (b) Concentration dependence of $\rho$$(T)$  for $x$=0.8, 0.2, 0.05, and 0.03 under a magnetic field of 2 T.
}
\label{fig:4}
\end{center}
\end{figure}

We confirmed that the superconductivity of the Pb phase is not responsible for the decrease below $T_{D}$ for at least $H$$\ge$1 T from the measurements with increasing thickness of the Pb phase,~\cite{rf:19} which is consistent with the results for PrPb$_{3}$~\cite{rf:20} and CePb$_{3}$.~\cite{rf:21} 
Figure 4(b)  shows $\rho$$(T)$ for $x=$0.8, 0.2, 0.05, and 0.03 at $H$=2 T. The nature of the scattering of the conduction electrons by the $\Gamma_{3}$ moments in the Pr dilute region seems to differ from that in the Pr concentrated region. For $x=$0.8 and 0.2, the fluctuation of $\Gamma_{3}$ moments is mainly suppressed due to the formation of RTLS, as we discussed in ref. 12. Therefore $T_{D}$ can be suppressed down to $\sim$1.5 K and the decrease below $T_{D}$ is a little.  On the other hand, for $x$=0.05 and 0.03,  the conduction electrons correlate with the fluctuating $\Gamma_{3}$ moments below $T_{D}$, leading to the marked decrease of the resistivity. If the fluctuation of the $\Gamma_{3}$ moments were suppressed due to the formation of the two-level system as in the case for $x=$0.8 and 0.2, $T_{D}$ and the magnitude of the decrease below $T_{D}$ should change monotonically with the Pr concentration. 

The Pr concentration dependence of the specific heat and resistivity is probably explained by considering the nearest-neighbor ion of a Pr site.  The probability that at least one of the six nearest-neighbor sites is occupied by a Pr ion (=1$-(1-x)^6$) is $\sim$50\% for $x$=0.1, suggesting that the distortion is still large. On the other hand, the probability for $x$=0.05 decreases down to $\sim$25\%, implying that the environment for Pr ions becomes fairly isotropic.  In the Pr dilute region, therefore,  the fluctuation may be suppressed via the correlation with the conduction electrons.

From the results of the resistivity,  it seems reasonable to suppose that the Kondo effect arising from the correlation between the $\Gamma _{3}$ moments and the conduction electrons gives rise to the scaling behavior in $\Delta$$C/T$.  However, the quantitative analysis of the  experimental results has not yet been carried out. $\rho$$(T)$ below $T_{D}$ shows a power law behavior $\rho$$(T)$$-$$\rho$$_{0}^{'}$$\propto$$BT^{n}$. The value of $n$ seems to vary with the temperature range, for example $n$=0.5 in 1.7$\le$$T$$\le$2.4 K with $B$= 0.28 ($\mu$$\Omega$cm/K$^{0.5}$) and $n$=1 in  0.6$\le$$T$$\le$1.5 K with $B$=0.10($\mu$$\Omega$cm/K)  for $x$=0.05 at $H$=2 T of Fig. 4(b).  Further experimental study and theoretical analysis taking account of the strain field and the magnetic field are needed to elucidate the features of the resistivity and the specific heat. 

In summary, we have studied the low-temperature properties of Pr$_{x}$La$_{1-x}$Pb$_{3}$ with the CEF ground state of a non-Kramers $\Gamma _{3}$ doublet for a wide concentration range of Pr ions. It was found that the temperature dependence of the specific heat and the electrical resistivity is drastically and qualitatively changed with the Pr concentration. For $x$$\le$0.05, the specific heat $\Delta$$C/T$ indicates an NFL behavior with an enhancement of $\Delta$$C/T$ below 1.5 K, which can be scaled with a characteristic temperature $T^{*}$ defined at each concentration $x$. The electrical resistivity in the corresponding temperature region shows a marked decrease deviating from a Fermi-liquid behavior $\rho$$(T)$$\propto$$T^{2}$. We suppose that the Kondo effect arising from  the correlation between the dilute $\Gamma _{3}$ moments and the conduction electrons gives rise to such anomalous behavior. 

The authors acknowledge valuable discussions with and helpful suggestions from Professor M. Koga and Professor H. Ishii. They also thank Professor Y. Makihara and Dr. K. Hashizume for technical help. This work was supported by the Grant-in-Aid for Scientific Research from the Ministry of Education, Culture, Sports, Science and Technology of Japan.


\begin{thebibliography}{99}
\bibitem{rf:1}  H. v. L\"{o}hneysen {\it et al.}:  Phys. Rev. Lett. {\bf 72} (1994) 3262 .
\bibitem{rf:2} B. Andraka {\it et al.}:  Phys. Rev. B {\bf 47} (1993) 3208 .
\bibitem{rf:3}  D. L. Cox: Phys. Rev. Lett. {\bf 59} (1987) 1240.
\bibitem{rf:4} I. Affleck {\it et al.}: Phys. Rev. B {\bf 48} (1993) 7297.
\bibitem{rf:5} C. L. Seaman {\it et al.}:  Phys. Rev. Lett. {\bf 67} (1991) 2882.
\bibitem{rf:6}  M. B. Maple {\it et al.}: J. Low Temp. Phys. {\bf 99} (1995) 223.
\bibitem{rf:7}  H. Amitsuka {\it et al.}:  J. Phys. Soc. Jpn. {\bf 63} (1994) 736.
\bibitem{rf:8} B. Andraka {\it et al.}:  Phys. Rev. Lett. {\bf 67} (1987) 2886.
\bibitem{rf:9}  H. Amitsuka  {\it et al.}: Physica B {\bf281\&282} (2000) 326.
\bibitem{rf:10}  A. Yatskar {\it et al.}:  Phys. Rev. Lett. {\bf 77} (1996) 3637.
\bibitem{rf:11}  E. D. Bauer {\it et al.}:  Phys. Rev. B {\bf 65} (2002) 100506(R).
\bibitem{rf:12}  T. Kawae {\it et al.}:   Phys. Rev. B {\bf 65} (2002) 012409.
\bibitem{rf:13}  W. Gross {\it et al.}: Z. Phys. B {\bf 37} (1980) 123.
\bibitem{rf:14}  E. Bucher {\it et al.}: Proc. Int. Conf. Low Temp. Phys:  Proc. Int. Conf. Low Temp. Phys. LT13, Vol{\bf 2} (1972) 322.
\bibitem{rf:15} H. Kusunose {\it et al.}: Phys. Rev. Lett. {\bf 76} (1996) 271.
\bibitem{rf:16}  P. D. Sacramento {\it et al.}:  Phys. Rev. B {\bf 43} (1991) 13294.
\bibitem{rf:17}  D. L. Cox  and A. Zawadowski: {\it Exotic Kondo Effects in Metal} (Taylor and Francis, London, 1999) p. 250.
\bibitem{rf:18} T. Kawae {\it et al.}: to be published in Physica B. 
\bibitem{rf:19} T. Kawae {\it et al.}: in preparation. 
\bibitem{rf:20}  Z. Kletowski {\it et al.}: J. Magn. Magn. Mater. {\bf 162} (1996) 277.
\bibitem{rf:21}  J. McDonough {\it et al.}: Phys. Rev. B {\bf 53} (1996) 14411.

\end{thebibliography}
\end{document}